\PassOptionsToPackage{unicode}{hyperref}
\PassOptionsToPackage{hyphens}{url}
\documentclass[
]{article}
\usepackage{amsmath,amssymb}
\usepackage{iftex}
\ifPDFTeX
  \usepackage[T1]{fontenc}
  \usepackage[utf8]{inputenc}
  \usepackage{textcomp} 
\else 
  \usepackage{unicode-math} 
  \defaultfontfeatures{Scale=MatchLowercase}
  \defaultfontfeatures[\rmfamily]{Ligatures=TeX,Scale=1}
\fi
\usepackage{lmodern}
\ifPDFTeX\else
\fi
\IfFileExists{upquote.sty}{\usepackage{upquote}}{}
\IfFileExists{microtype.sty}{
  \usepackage[]{microtype}
  \UseMicrotypeSet[protrusion]{basicmath} 
}{}
\makeatletter
\@ifundefined{KOMAClassName}{
  \IfFileExists{parskip.sty}{%
    \usepackage{parskip}
  }{
    \setlength{\parindent}{0pt}
    \setlength{\parskip}{6pt plus 2pt minus 1pt}}
}{
  \KOMAoptions{parskip=half}}
\makeatother
\usepackage{xcolor}
\usepackage[margin=1in]{geometry}
\usepackage{longtable,booktabs,array}
\usepackage{calc} 
\usepackage{etoolbox}
\makeatletter
\patchcmd\longtable{\par}{\if@noskipsec\mbox{}\fi\par}{}{}
\makeatother
\IfFileExists{footnotehyper.sty}{\usepackage{footnotehyper}}{\usepackage{footnote}}
\makesavenoteenv{longtable}
\usepackage{graphicx}
\makeatletter
\def\maxwidth{\ifdim\Gin@nat@width>\linewidth\linewidth\else\Gin@nat@width\fi}
\def\maxheight{\ifdim\Gin@nat@height>\textheight\textheight\else\Gin@nat@height\fi}
\makeatother
\setkeys{Gin}{width=\maxwidth,height=\maxheight,keepaspectratio}
\makeatletter
\def\fps@figure{htbp}
\makeatother
\NewDocumentCommand\citeproctext{}{}

\makeatletter
 \let\@cite@ofmt\@firstofone
 \def\@biblabel#1{}
 \def\@cite#1#2{{#1\if@tempswa , #2\fi}}
\makeatother
\newlength{\cslhangindent}
\newlength{\csllabelwidth}
\newenvironment{CSLReferences}[2] 
 {\begin{list}{}{%
  \setlength{\itemindent}{0pt}
  \setlength{\leftmargin}{0pt}
  \setlength{\parsep}{0pt}
  \ifodd #1
   \setlength{\leftmargin}{\cslhangindent}
   \setlength{\itemindent}{-1\cslhangindent}
  \fi
  \setlength{\itemsep}{#2\baselineskip}}}
 {\end{list}}
\usepackage{calc}

\ifLuaTeX
  \usepackage{selnolig}  
\fi
\usepackage{bookmark}
\IfFileExists{xurl.sty}{\usepackage{xurl}}{} 
\hypersetup{
  pdftitle={OptimOTU: Taxonomically aware OTU clustering with optimized thresholds and a bioinformatics workflow for metabarcoding data},
  pdfauthor={Brendan Furneaux1; Sten Anslan1; Panu Somervuo2; Jenni Hultman3,4; Nerea Abrego1; Tomas Roslin2,5; Otso Ovaskainen1},
  hidelinks,
  pdfcreator={LaTeX via pandoc}}

\title{OptimOTU: Taxonomically aware OTU clustering with optimized thresholds and a bioinformatics workflow for metabarcoding data}
\author{Brendan Furneaux\textsuperscript{1} \and Sten Anslan\textsuperscript{1} \and Panu Somervuo\textsuperscript{2} \and Jenni Hultman\textsuperscript{3,4} \and Nerea Abrego\textsuperscript{1} \and Tomas Roslin\textsuperscript{2,5} \and Otso Ovaskainen\textsuperscript{1}}
\date{\textsuperscript{1}Department of Biological and Environmental Science, University of Jyväskylä, Jyväskylä, Finland\\
\textsuperscript{2}Faculty of Biological and Environmental Sciences, University of Helsinki, Helsinki, Finland\\
\textsuperscript{3}Department of Microbiology, University of Helsinki, Helsinki, Finland\\
\textsuperscript{4}Natural Resources Institue Finland (Luke), Helsinki, Finland\\
\textsuperscript{5}Department of Ecology, Swedish University of Agricultural Sciences, Uppsala, Sweden}

\begin{document}
\maketitle

\section*{Abstract}\label{abstract}
\addcontentsline{toc}{section}{Abstract}

To turn environmentally derived metabarcoding data into community matrices for ecological analysis, sequences must first be clustered into operational taxonomic units (OTUs).
This task is particularly complex for data including large numbers of taxa with incomplete reference libraries.
OptimOTU offers a taxonomically aware approach to OTU clustering.
It uses a set of taxonomically identified reference sequences to choose optimal genetic distance thresholds for grouping each ancestor taxon into clusters which most closely match its descendant taxa.
Then, query sequences are clustered according to preliminary taxonomic identifications and the optimized thresholds for their ancestor taxon.
The process follows the taxonomic hierarchy, resulting in a full taxonomic classification of all the query sequences into named taxonomic groups as well as placeholder ``pseudotaxa'' which accommodate the sequences that could not be classified to a named taxon at the corresponding rank.
The OptimOTU clustering algorithm is implemented as an R package, with computationally intensive steps implemented in C++ for speed, and incorporating open-source libraries for pairwise sequence alignment.
Distances may also be calculated externally, and may be read from a UNIX pipe, allowing clustering of large datasets where the full distance matrix would be inconveniently large to store in memory.
The OptimOTU bioinformatics pipeline includes a full workflow for paired-end Illumina sequencing data that incorporates quality filtering, denoising, artifact removal, taxonomic classification, and OTU clustering with OptimOTU.
The OptimOTU pipeline is developed for use on high performance computing clusters, and scales to datasets with millions of reads per sample, and tens of thousands of samples.

\section{Introduction}\label{introduction}

Clustering of environmentally derived marker gene sequences into operational taxonomic units (OTUs) prior to downstream analysis is a common step in molecular ecology workflows.
The aim is to group sequences that are derived from the same species or strain, but differ due to errors during amplification or sequencing, or due to true genetic variation between gene copies in the same genome or between individuals.

Traditional agglomerative and greedy clustering algorithms treat these sources of variation in the same way, clustering sequences into OTUs based on a single global threshold (Dondoshansky and Wolf 2000; W. Li and Godzik 2006; Edgar 2010; Mahé et al. 2015; Ratnasingham and Hebert 2013).
In this case, the threshold should be chosen to be larger than typical error rates for the chosen sequencing workflow, and also larger than the intraspecies genetic variation present within the species of interest.
At the same time, it should be smaller than the level of genetic variation between species.

As an alternative approach, denoising algorithms such as Deblur (Amir et al. 2017), DADA2 (Callahan et al. 2016), and UNOISE2 (Edgar 2016b) attempt to separate true biological variation from sequencing errors.
By incorporating information on sequence abundance (and quality scores, in the case of DADA2), these methods pinpoint sequence variants unlikely to derive from sequencing errors, which are known variably as amplicon sequence variants (ASVs), sub-OTUs (sOTUs), or zero-diameter OTUS (zOTUs).

Denoising may successfully recover true biological variation even below the sequencing error rate, and in some cases ASVs have been used directly in ecological analysis (Callahan, McMurdie, and Holmes 2017).
However, this ability to distinguish very similar variants also frequently separates sequences that are derived from the same species or strain (Kauserud 2023), and so ASVs are often subjected to a second clustering step to group sequences using a biologically motivated threshold.

In both direct clustering and post-denoising clustering, it is common to choose a single threshold, often 97\% similarity, to form OTUs from the sequences, even across large and diverse groups such as invertebrates (e.g., Watts et al. 2019), fungi (e.g., Tedersoo et al. 2021), or microbial eukaryotes (e.g., Jamy et al. 2020).
However, levels of genetic variation within species are known to vary between taxonomic groups as well as between marker genes (Nilsson et al. 2008; Pentinsaari et al. 2016), so these thresholds may result in over- or under-clustering, and most often both within the same dataset.

An additional challenge is that environmental samples of highly diverse organism groups typically include a large number of taxa which are not represented in reference libraries.
Thus, a large proportion of the sequences actually derive from taxa which cannot be assigned to a cluster present in the reference set, and for which levels of genetic variation cannot be reliably established.

\begin{figure}
\centering
\includegraphics{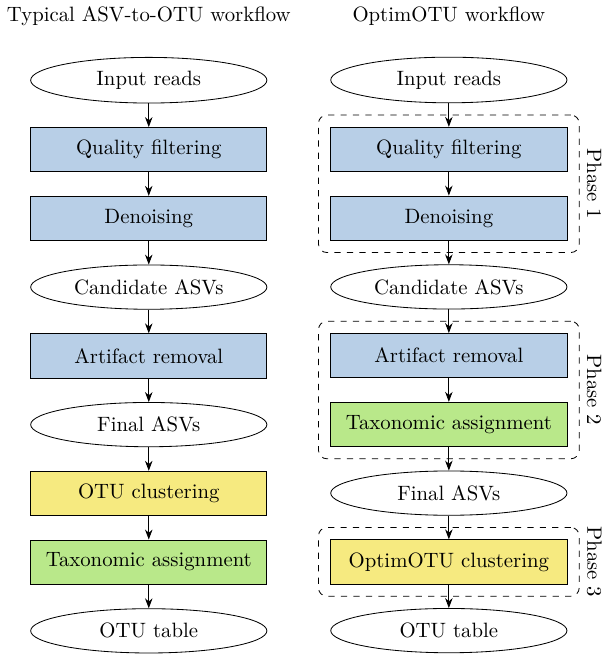}
\caption{\label{fig:pipeline-overview}Overview of the OptimOTU workflow
A typical ASV-to-OTU workflow is shown on the left for comparison.
The three phases of the OptimOTU pipeline are shown in more detail in Fig \ref{fig:pipeline-phase1}, Fig \ref{fig:pipeline-phase2}, and Fig \ref{fig:pipeline-phase3}.}
\end{figure}

Here we present the OptimOTU clustering algorithm, a taxonomically aware approach to OTU clustering, in which optimized thresholds are used for different taxonomic groups.
We also present the OptimOTU bioinformatics pipeline, which incorporates the OptimOTU clustering algorithm into a full workflow for processing paired-end Illumina sequencing data into OTUs (Fig \ref{fig:pipeline-overview}).

\section{Implementation}\label{implementation}

The OptimOTU clustering algorithm has three phases: threshold optimization, preliminary taxonomic identification, and clustering.
The threshold optimization and clustering phases are implemented in the R package \texttt{optimotu}, available at \url{https://github.com/brendanf/optimotu}, while preliminary taxonomic identification can be performed by a variety of third-party algorithms.
Threshold optimization is usually the most computationally intensive step, but it does not need to be repeated for every study if a set of optimized thresholds for the taxonomic groups of interest has already been determined.

The threshold optimization procedure is similar to the one used by dnabarcoder (Vu, Nilsson, and Verkley 2022).
It is performed in the OptimOTU clustering algorithm by the function \texttt{optimize\_thresholds()}, which takes as input a set of reference sequences and their taxonomic identifications.
These can be obtained from a curated global database such as BOLD data release packages (Ratnasingham and Hebert 2013), UNITE (Abarenkov et al. 2010), PR2 (Guillou et al. 2013), or Eukaryome (Tedersoo et al. 2024), or from a local database of sequences.
It is also possible to use a subset of the focal sequences that has been identified with high confidence by algorithms such as the RDP naive Bayesian classifier (Wang et al. 2007), SINTAX (Edgar 2016a), PROTAX (Somervuo et al. 2016; Abarenkov et al. 2018; Roslin et al. 2022), or similar methods.
This was, for example, the approach used by Ovaskainen et al. (2024).

\begin{figure}
\centering
\includegraphics{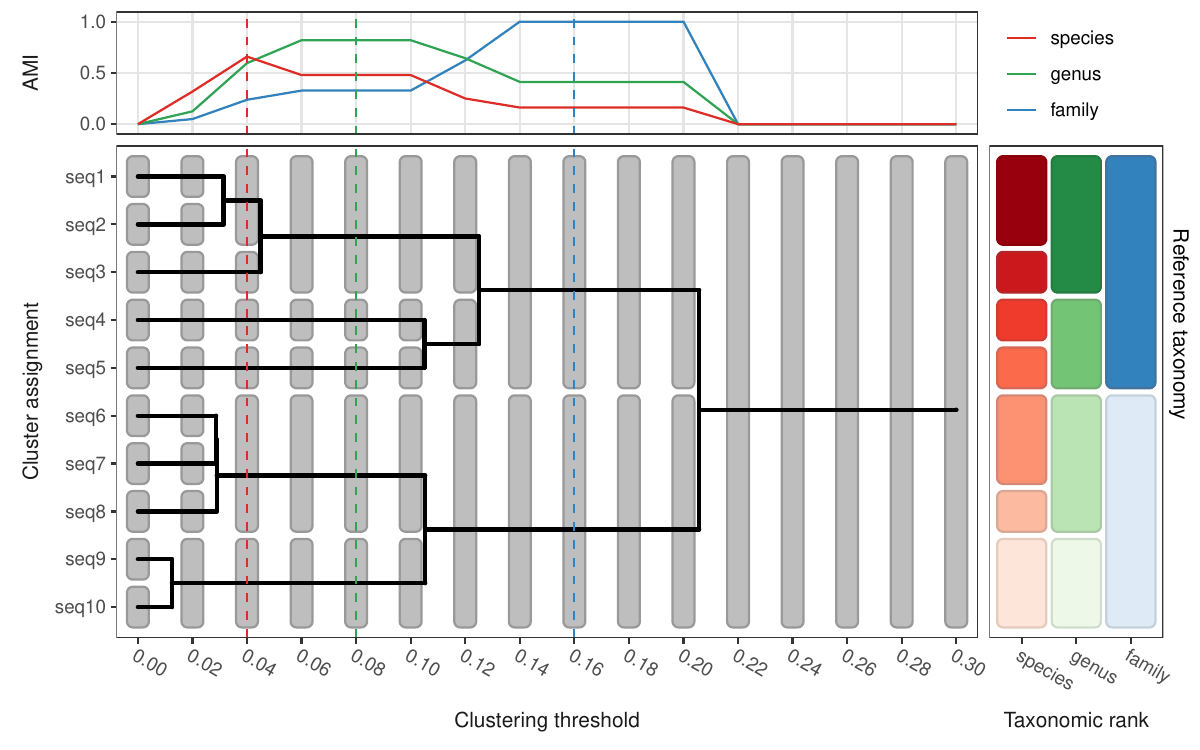}
\caption{\label{fig:cluster-optimization}OptimOTU threshold optimization procedure.
Here we show a constructed example for optimization of clustering thresholds at the ranks of family, genus, and species, within a single taxonomic order.
The taxonomic identity for the ten reference sequences are shown as colored bars on the right.
The full hierarchical clustering tree is shown in the left, with clustering partitions using different thresholds shown as gray bars.
The adjusted mutual information (AMI) score calculated for the clustering at each threshold in comparison to the reference taxonomy at the three ranks is shown at top.
Dashed vertical lines indicate thresholds which optimize AMI for each rank, with ties resolved by selecting the median threshold and rounding down.}
\end{figure}

The sequences are then clustered by single-linkage hierarchical agglomerative clustering, resulting in a tree structure which can be cut at any level to produce a partition of the reference sequences into clusters.
Such partitions are calculated for a range of test thresholds, for example 0.0 to 0.4 in steps of 0.001.
The resulting clusters are then compared to the taxonomic identifications, to determine the threshold that produces clusters that most closely match the taxonomy at each rank (Fig \ref{fig:cluster-optimization}).
This process is conceptually repeated for each taxon in the reference set which has enough representative sequences (default: 10) and enough descendant taxa (default: 5) to be informative.
In practice, clustering is performed concurrently for all relevant taxa, with only a single pass through the sequences.

Once the optimized thresholds have been determined, they can be used to cluster query sequences.
It is worth noting that OptimOTU inverts the typical workflow in which sequences are clustered first and then taxonomically identified, and instead uses preliminary taxonomic identification first, followed by clustering informed by the identifications (Fig \ref{fig:pipeline-overview}).
The taxonomic identification can be performed by any method with a stopping condition, meaning that taxonomic identification is truncated to higher ranks if the confidence is low.
The set of applicable methods includes any of the wide range of algorithms which produce confidence scores for classifications at each rank.
This includes the popular k-mer profile naive Bayesian classifier first implemented by the RDP classifier (Wang et al. 2007) but also available in software such as Mothur (Schloss et al. 2009), QIIME 2 (Bokulich et al. 2018) and DADA2 (Callahan et al. 2016).
Rankwise confidence scores are also produced by SINTAX (Edgar 2016a), PROTAX (Somervuo et al. 2016; Abarenkov et al. 2018; Roslin et al. 2022), IDTAXA (Murali, Bhargava, and Wright 2018), Gappa (Czech, Barbera, and Stamatakis 2020), BayesANT (Zito, Rigon, and Dunson 2023), and MycoAI (Romeijn, Bernatavicius, and Vu 2024).
Additionally, algorithms based on last common ancestor (LCA) consensus such as implemented in CREST (Lanzén et al. 2012), USEARCH, (Edgar 2010), VSEARCH (Rognes et al. 2016), and QIIME 2 (Bokulich et al. 2018) may be suitable.
Taxonomic identification prior to clustering is used to choose thresholds in the clustering process, and also to constrain the clustering.
These constraints both ensure that the resulting clustering is congruent with the taxonomic identifications, and also serve to split the queries into taxonomically defined groups, increasing the computational efficiency of the clustering process.

\begin{figure}
\centering
\includegraphics{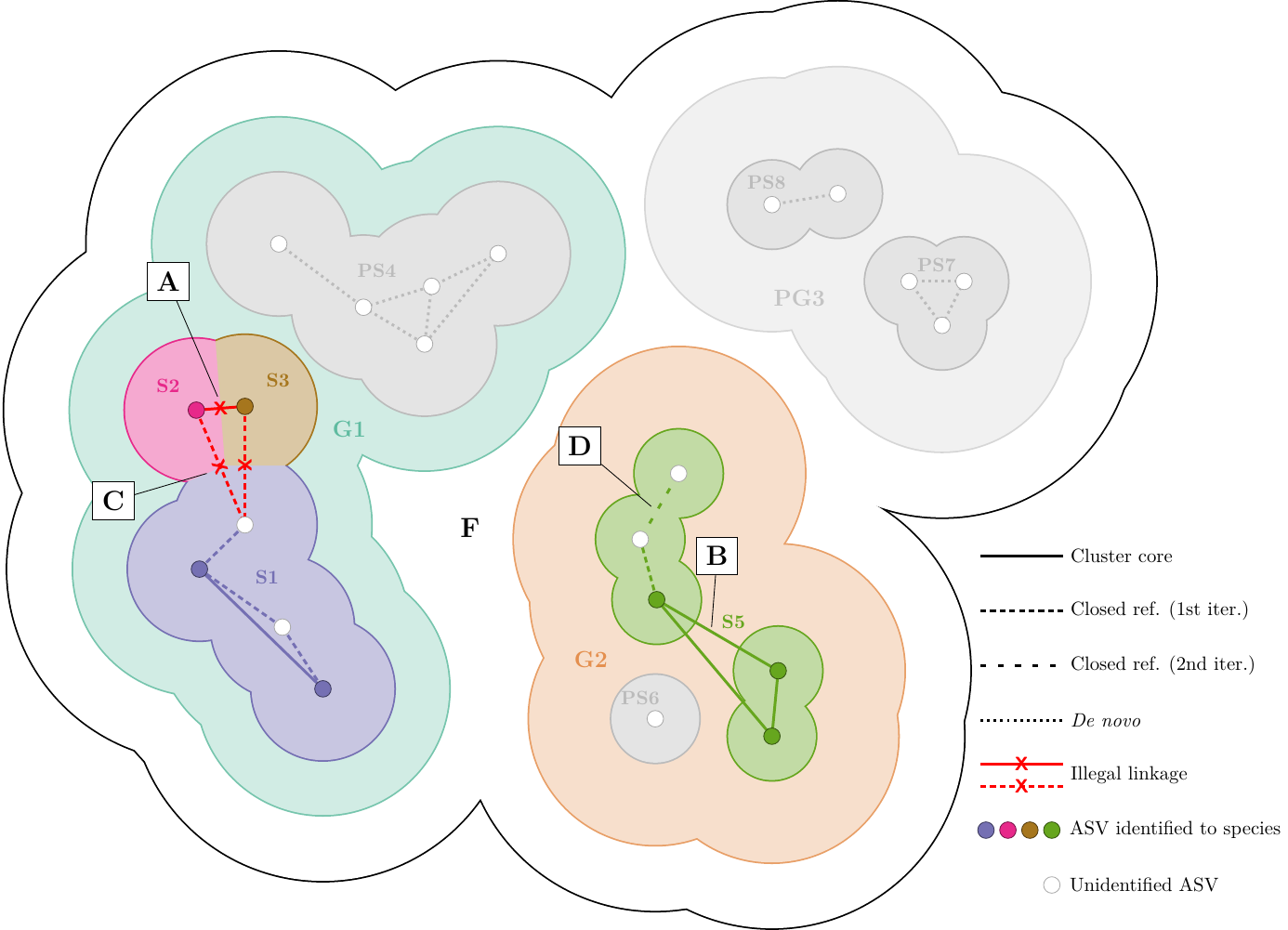}
\caption{\label{fig:clustering}OptimOTU clustering algorithm.
The same procedure is used at all ranks, but here is illustrated at the species level.
In this constructed example, ASVs belonging to family F have already been clustered into two named genera G1 and G2 and one pseudogenus PG3 during previous iterations.
Clustering thresholds are depicted as circular boundaries around ASVs, such that when the circles overlap, the ASVs are more similar than the threshold.
In this example, an optimized species-level clustering threshold has been determined for G1 based on reference data, but not for G2.
Thus, ASVs in G2 are clustered using a species-level threshold inherited from F.
Because it is not possible to optimize clustering within a pseudotaxon, ASVs in PG3 are also clustered using the threshold inherited from F.
In the first stage, cluster cores are formed from ASVs which have been identified to species (colored dots), regardless of genetic distance.
Thus the ASVs belonging to species S2 and S3 are not linked (A), despite being within the species-level threshold, while all three ASVs belonging to species S5 are linked (B), despite one of them being outside the species-level threshold from either of the others.
Next, closed-reference clustering links unidentified ASVs (white dots) to the nearest cluster core if they are within the species-level threshold.
Note that this does not merge species S1 with S2 or S3 (C), despite the presence of an ASV which is within the threshold of all three cluster cores; it is only merged with the closest cluster core.
The closed-reference clustering step is iterated as long as new matches are found, allowing an additional ASV to be linked to species S5 (D).
Finally, ASVs which have not been linked to any cluster core are \emph{de novo} clustered, forming pseudospecies ps4, ps6, ps7, and ps8.}
\end{figure}

The clustering process is then performed using the \texttt{optimotu()} function, which proceeds in a hierarchical manner through the taxonomic ranks, starting with the rank below the root.
Within each rank, the sequences are grouped by their cluster identity at the rank above and processed separately (Fig \ref{fig:clustering}).
At each rank, the clustering proceeds in three stages.
First, cluster cores are formed by grouping sequences which have the same taxonomic identifications at the current rank.
Second, sequences which could not be identified at the current rank are assigned to cluster cores by closed-reference clustering, using the optimized threshold for the closest enclosing taxon for which a threshold was determined.
The closed-reference clustering is performed by pairwise comparisons between the unassigned sequences and all sequences in the cluster cores, and if the distance is less than the threshold, the sequence is added to the cluster.
If the sequence is closer than the threshold to sequences from multiple cluster cores, then it is assigned to the one with the closest match.
If there is a tie between multiple sequences in different cluster cores, then the new sequence is assigned at random to the cluster core of one of the tied sequences.

The closed reference clustering phase is performed iteratively, with the sequences newly assigned to cluster cores in the previous iteration being used as the target sequences in the next iteration, until no new matches are found.
In this way, the resulting partition approximates single-linkage clustering.
However, clusters which would be joined in full single-linkage may be separated if they contain sequences with different taxonomic identifications.
In the third and final step, sequences which are still unassigned at the chosen rank are \emph{de novo} single-linkage clustered with the optimized threshold.
The clusters from the \emph{de novo} clustering are given placeholder taxonomic names of the form ``pseudo\{rank\}\_\{number\}``, where \{rank\} is the rank of the cluster and \{number\} is a unique integer identifier.

When the clustering is complete, the result is a full taxonomic classification for each sequence, including named taxa and pseudotaxa.
Clusters at the lowest rank, typically species, are then used as OTUs for further analysis.

\subsection{Pairwise distance calculation algorithms}\label{pairwise-distance-calculation-algorithms}

The OptimOTU package can use various methods for calculating pairwise distances between sequences.
Three methods are implemented internally in the R package and can be selected using the option \texttt{dist\_config} to \texttt{optimize\_thresholds()} or \texttt{optimotu()}.

The first method is a simple Hamming distance, which counts the number of differences between two sequences.
Although this is by far the fastest method, it is only suitable if both the query and reference sequences are already globally aligned.
This is the recommended method for use with protein-coding sequences such as COI, and may also be applied to the ribosomal small subunit (SSU/12S/16S/18S), 5.8S, and large subunit (16S/23S/25S/26S/28S) RNA.
It should not be used for sequences containing introns, spacers such as ITS, or other regions where multiple alignment across broad taxonomic ranges is not reliable.
The implementation is adapted from ProtaxA (Roslin et al. 2022) which uses a 4-bit one-hot encoding of DNA sequences and the bit-count instruction of modern processors to accelerate the calculation.
It is activated using \texttt{dist\_config\ =\ dist\_hamming(min\_overlap,\ ignore\_gaps)}.
Here, the two options control handling of gap characters.
The \texttt{min\_overlap} defines a minimum number of sites which contain a nucleotide in both sequences being compared.
Low pairwise alignment overlap is expected to be rare in a global multiple sequence alignment of genuine metabarcoding reads of ``alignable'' markers such as 16S or COI, which should consist of full-length reads of the homologous region between two primers.
However, low overlap may still occur in the case of artefactual sequences, and may also be more common in reference databases, where sequences may come from different primer sets or have been trimmed to different lengths.
The \texttt{ignore\_gaps} option allows the user to specify whether sites where a gap in one sequence is aligned to a nucleotide in the other sequence should be ignored in the distance calculation or counted as a mismatch.
End gaps and gap-gap sites are always ignored.

The second method uses the open-source C++ library Edlib (Šošić and Šikić 2017) to calculate edit (Levenshtein) distance between two sequences using Needleman-Wunsch dynamic programming.
Edlib also allows the use of banding to limit the search space, and early stopping when the alignment score exceeds a predefined threshold.
This is the fastest internal method for calculating distances between unaligned sequences when the distances are large.
However, the pairwise alignments may not be as biologically relevant as those using more complex gap-affine scoring schemes.
The Edlib method is activated using \texttt{dist\_config\ =\ dist\_edlib()}.

The third method uses wavefront alignment (WFA) with the open-source C++ library WFA2 (Marco-Sola et al. 2021).
WFA is especially efficient for very similar sequences.
In addition to the edit distance, it can also calculate gap-linear, gap-affine, and dual-cost gap-affine alignments.
Note, however, that although the alignment may be calculated with various scoring functions, the distance used by the OptimOTU package is always equivalent to the Hamming distance between the aligned sequences (not including end gaps) divided by the alignment length, as used by, e.g., USEARCH (Edgar 2010).
Like Edlib, WFA2 support banding and early stopping.
It is activated by setting \texttt{dist\_config\ =\ dist\_wfa2(match,\ mismatch,\ gap\_open,\ gap\_extend,\ gap\_open2,\ gap\_extend2)}, where the arguments are the scores for the dual-cost gap-affine scoring function.
When the supplied scores are equivalent to a simpler scoring function (as in the default case, which is edit distance) the alignment is calculated using that simpler function.
WFA2 in edit-distance mode is the default for the OptimOTU package.

Finally, in the fourth method distances are calculated externally and supplied as a distance matrix in three-column (query, reference, distance) format, which can be read from a file or UNIX pipe, allowing the use of arbitrary external tools.
This alternative is activated by setting \texttt{dist\_config\ =\ dist\_file(filename,\ by\_name)}, where \texttt{filename} is the name of the file or pipe, and \texttt{by\_name} is a logical value indicating whether the sequences are identified by name or by 0-based index.
Although an external distance matrix can be used by both \texttt{optimize\_thresholds()} and \texttt{optimotu()}, the distances must be read more than once in \texttt{optimotu()}, making it incompatible with piped input.
However, an additional method is implemented which runs USEARCH (Edgar 2010) to calculate sparse distance matrices via its \texttt{calc\_distmx} command for the optimization and \emph{de novo} clustering stages, and \texttt{usearch\_global} for the closed-reference clustering stage.
This method is activated by setting \texttt{dist\_config\ =\ dist\_usearch(usearch,\ usearch\_ncpu)}, where \texttt{usearch} is the path to a suitable USEARCH executable, and \texttt{usearch\_ncpu} is the number of threads to request from each invocation of USEARCH.
Note that these threads are independent of the threads used by the OptimOTU clustering algorithm, and the total number of threads used by \texttt{dist\_usearch()} is the sum of \texttt{usearch\_ncpu} and the \texttt{threads} argument to the chosen parallelization strategy (see below).
Unlike the other distance calculation methods, \texttt{dist\_usearch()} configures USEARCH to use shared k-mer heuristics to choose candidate pairs for alignment in order to avoid calculating the full distance matrix.
This means it is typically the fastest option for markers such as ITS where global multiple sequence alignment is not feasible, but some pairwise distances which would alter the clustering result may be missing.

\subsection{Clustering algorithms}\label{clustering-algorithms}

The OptimOTU package implements several single-linkage clustering algorithms, of which SLINK and a novel tree-based algorithm are recommended for use.

The SLINK algorithm (Sibson 1973) operates at the theoretically optimal complexities of \(O(n^2)\) time and \(O(n)\) space, where \(n\) is the number of sequences.
Unfortunately, SLINK is an inherently serial algorithm which must process the distances in a predefined order.
Therefore, it is not possible to trivially parallelize (but see Nolet et al. 2023).
The OptimOTU implementation of SLINK allows clustering based on an incomplete distance matrix, provided that the elements which are present are in the correct order.
When it detects one or more ``skipped'' distances in its input, it proceeds as if the missing elements were at infinite distance.
It also allows calculation, before each new pairwise distance is processed, of the maximum distance which would need to be obtained in order to cause an update.
This is used to adaptively set the band width and maximum alignment score for Edlib and WFA2, in order to reduce the number of unnecessary calculations.
The SLINK algorithm can be activated using the option \texttt{clust\_config\ =\ clust\_slink()} to \texttt{optimize\_thresholds()} or \texttt{optimotu()}.

The tree-based algorithm maintains a full hierarchical clustering tree at all stages of clustering, which is updated as each new pairwise distance is processed.
Although this algorithm has a higher time complexity than SLINK, it can process the distances in any order.
Like SLINK, the tree-based algorithm can calculate the maximum relevant distance for a given sequence pair, and use this distance to set the band width and maximum alignment score for Edlib or WFA2.
Because it is compatible with out-of-order distance matrices, the tree-based algorithm is the default clustering algorithm in the OptimOTU package, but it can be activated or configured using the option \texttt{clust\_config\ =\ clust\_tree()} to \texttt{optimize\_thresholds()} or \texttt{optimotu()}.

The two clustering algorithms should always produce identical partitions when run on the same distances.
The test suite for the OptimOTU package tests the clustering algorithms with all applicable parallelization options for consistency with each other and the base R \texttt{hclust()} command.
Additionally, internal consistency checks can be activated by setting the optional argument \texttt{test\ =\ 1} or \texttt{test\ =\ 2} in \texttt{clust\_tree()}.
Since these checks are computationally expensive, they are not recommended for routine use.
However, they offer useful checks for bugs in the tree-based algorithm.

\subsection{Parallelization}\label{parallelization}

There are three available parallelization options, which are configured by setting the \texttt{parallel\_config} argument in \texttt{optimize\_thresholds()} or \texttt{optimotu()}.

The first method relies on the ability of the tree-based algorithm to process pairwise distances in any order, and to calculate the maximum relevant distance for a given pair before processing it.
Because the majority of pairwise distances do not trigger updates, multiple parallel threads can cluster (and calculate distances) using two-stage locking.
In the first step, the tree is non-exclusively locked for reading to calculate the maximum relevant distance for a given pair and the distance is calculated.
Then, in the second step the tree is exclusively locked for writing only if a relevant distance is found.
This is referred to as the ``concurrent'' option, selected using \texttt{parallel\_config\ =\ parallel\_concurrent(threads)}, where \texttt{threads} is the number of parallel threads to use for updating the tree.

It is also possible to merge two trees using only \(O(n)\) updates, which allows an alternate parallelization technique, where multiple tree structures are processed in parallel and then merged at the end.
This technique can also be used to cluster different subsets of the distance matrix in parallel using SLINK, and then merge the results using the tree-based algorithm.
This method is referred to as the ``merge'' option, accessed by setting the option \texttt{parallel\_config\ =\ parallel\_merge(threads)}, where \texttt{threads} is the number of tree structures, each updated by its own thread.
This is the only parallelization algorithm which can be used with SLINK if \texttt{threads} \(> 1\).

Finally, the two parallelization techniques can be combined, with multiple tree structures each updated by multiple parallel threads, and then merged at the end.
This is referred to as the ``hierarchical'' option, accessed by setting the option \texttt{parallel\_config\ =\ parallel\_hierarchical(threads,\ shards)}, where \texttt{threads} is the total number of threads and \texttt{shards} is the number of tree structures.

\subsection{Measures of clustering quality}\label{measures-of-clustering-quality}

To compare the quality of a clustering partition relative to a reference taxonomy, the OptimOTU package offers several measures.
The first group of measures are based on pair counting, where deciding whether each pair of sequences belong to the same cluster is treated as a binary classification problem.
Thus, a \(2\times2\) confusion matrix can be constructed.
If the sequence pair is grouped together in both the reference taxonomy and in the clustering partition, it is a true positive (TP), if they are in different clusters in both, it is a true negative (TN), if they are in the same cluster in the reference but different clusters in the partition, it is a false negative (FN), and if they are in different clusters in the reference but the same cluster in the partition, it is a false positive (FP).
Various measures can be calculated from the confusion matrix, of which the Rand index (RI; Rand 1971), adjusted Rand index (ARI; Hubert and Arabie 1985), Fowlkes-Mallows index (FMI; Fowlkes and Mallows 1983), and Matthews correlation coefficient (MCC; Matthews 1975) are implemented in the OptimOTU package.
Of these, the RI is not corrected for chance, so it tends to give higher scores to small clustering thresholds, i.e., clusterings with many small clusters.
For this reason, it is not recommended.
Because the calculation of the confusion matrix is the most computationally expensive part of calculating these measures, it is more efficient to calculate the confusion matrix once and then calculate multiple measures from it if multiple measures are desired.
This approach is supported in the OptimOTU package by the function \texttt{confusion\_matrix()},
which is parallelized, and a function for each index which takes the confusion matrix as input.

The second group of measures are based in information theory, and include the mutual information (MI; Strehl and Ghosh 2002) and adjusted mutual information (AMI; Vinh, Epps, and Bailey 2010).
These measures consider the cross-entropy between clusters in the test partition and the reference taxonomy.
Because the MI is not corrected for chance and has no consistent upper bound, it is recommended to always use AMI if an information theoretic measure is desired.

The third group is set-matching measures, where a single measure, the weighted multiclass F-measure (FM; Steinbach 2000), is implemented.
Unlike the previous measures, the FM considers only the best-matching cluster for each taxon in the reference taxonomy, so it is not sensitive to the number or makeup of any ``extra'' clusters in the test partition.
FM is the measure used for optimization in dnabarcoder (Vu, Nilsson, and Verkley 2022).

The default behavior of \texttt{optimize\_thresholds()} is to calculate thresholds which optimize each of the supported measures.
The user may then choose which measure to use for clustering by setting the \texttt{measure} argument to \texttt{optimotu()}.
Alternatively, one or more measures may be selected by setting the \texttt{measures} argument to \texttt{optimize\_thresholds()}.

\subsection{The OptimOTU pipeline}\label{the-optimotu-pipeline}

The OptimOTU pipeline is a full bioinformatics pipeline built around the OptimOTU clustering algorithm.
It is used to process demultipexed paired-end reads into a species\textsubscript{x}sample occurrence matrix and taxonomy for each OTU.
The pipeline is implemented using the \texttt{targets} workflow management package (Landau 2021), and is designed to be run across multiple nodes of a high-performance computing cluster using the \texttt{crew.cluster} backend (Landau, Levin, and Furneaux 2024), enabling it to process very large metabarcoding projects.
For smaller projects, it can also be run on a laptop or desktop computer.
The \texttt{targets} package allows the pipeline to skip already completed steps when re-run, which is useful both for debugging, and for ``on-line'' processing of new samples as they are sequenced.
Intermediate results (``targets'') are stored either as standard file types if they are input to external tools, or using the high-performance QS format for arbitrary R objects (Ching 2024), or FST for tabular data (Klik 2022).
Individual steps in the pipeline are implemented as R functions in the package \texttt{optimotu.pipeline}, available at \url{https://github.com/brendanf/optimotu.pipeline}, while the \texttt{targets} pipeline itself is implemented in a separate repository, \url{https://github.com/brendanf/optimotu_targets}.
An alternate containerized implementation of the same processing steps for execution on a single computer using the PipeCraft 2 graphical interface (Anslan et al. 2017, https://pipecraft2--manual.rtfd.io) is in progress.

The pipeline was originally developed for fungal ITS2 sequences in the Global Spore Sampling Project (Ovaskainen et al. 2024), but has also been specifically adapted for use with metazoan COI sequences.
Example configurations for these two marker genes are included in the pipeline codebase, and can be used as templates for other marker genes.

In brief, the workflow consists of primer removal, read quality filtering, denoising, removal of tag-switches, \emph{de novo} and reference-based chimera removal, probabilistic taxonomic assignment, and finally taxonomically-guided hierarchical clustering.
It also includes a number of optional steps which may only be relevant for certain marker genes, taxonomic groups, or laboratory workflows.
These include detection of artificial ``spike-in'' sequences, detecting and removing outgroup sequences, alignment to a profile hidden Markov model (HMM) or covariance model (CM), detection of nuclear mitochondrial pseudogenes (NuMts), or assignment of sequences to ecological guilds or functional groups.
Configuration of these optional steps, as well as settings for the standard steps, is performed using a human-readable YAML configuration file.

The pipeline is organized broadly into three phases, which are distinguished by the unit of parallelization (Fig \ref{fig:pipeline-overview}).
In the first phase, most operations are performed on individual samples, although some steps share information between all samples within a sequencing run, and all steps are performed in batches of no more than 96 samples.
This phase ends with the production of a master list of candidate ASVs and a table of read counts for each candidate ASV in each sample.
In the second phase, operations are performed on the master list of candidate ASVs, which is split into batches for parallel processing.
Operations at this stage primarily consist of various methods to filter the ASVs to remove artifacts, producing a final ASV list, as well as preliminary taxonomic identification.
In the third phase, the final ASV list is clustered into OTUs using the OptimOTU clustering algorithm.
After the main processing steps are complete, optional post-processing steps may be performed, and the results are written to disk.

Although many of the steps in the pipeline are performed in R using various freely available packages, several external command-line software tools are also used.
The R version, R packages, and external tools are enumerated along with recommended version numbers in an included Conda environment file (conda contributors, n.d.) which can be used to install the dependencies automatically, or can be read by the user to install the required software manually.

\subsubsection{Input files}\label{input-files}

The primary input to the pipeline is a set of paired-end FASTQ files, two (R1 and R2) for each sample.
Gzipped files are supported.
In the simplest cases, the sample names and read pairings are encoded in the file names, and can be automatically inferred by the pipeline.
In these cases, the reads should be placed in a directory structure within \texttt{sequences/01\_raw/} in the project directory, with each sequencing run in a separate subdirectory.
Multiple samples with identical names are supported as long as they are in different sequencing runs.
Additional levels of directory structure or non-FASTQ files are ignored.

Alternatively, the sample names, sequencing run identifiers, and file paths, as well as additional sample-specific information, can be specified in a TSV file.
This file is typically placed in a \texttt{metadata/} directory in the project directory, and is specified in the configuration file with the option \texttt{custom\_sample\_table}.
This option may be useful if the project is set up to use raw reads from a public repository such as the Sequence Read Archive (SRA) or European Nucleotide Archive (ENA).

\begin{figure}
\centering
\includegraphics{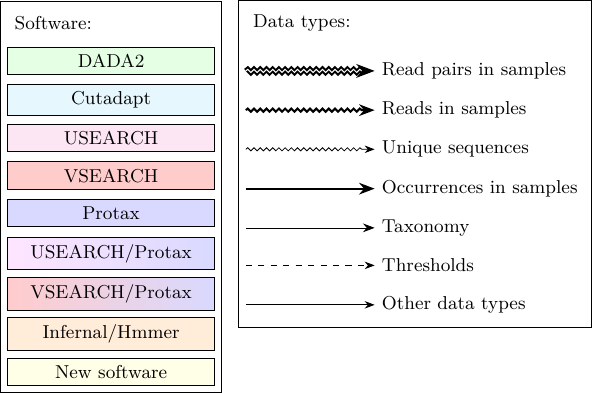}
\caption{\label{fig:pipeline-key}Key to the symbols used in OptimOTU pipeline flowcharts.}
\end{figure}

\begin{figure}
\centering
\includegraphics{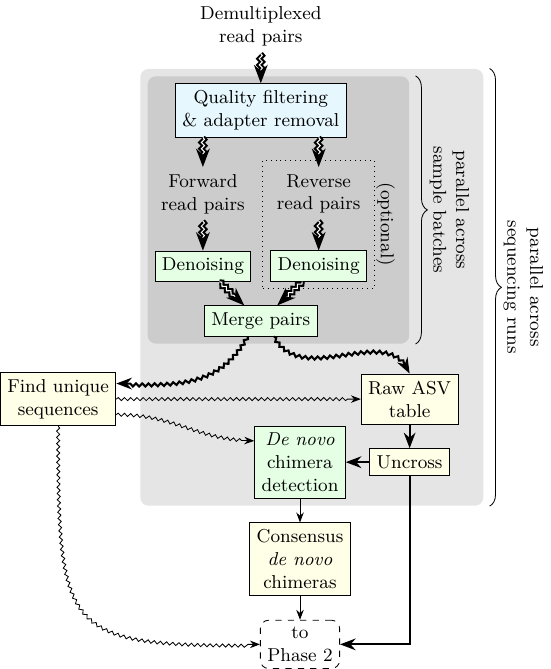}
\caption{\label{fig:pipeline-phase1}Phase 1 of the OptimOTU pipeline.}
\end{figure}

\subsubsection{Phase 1}\label{phase-1}

The first phase of the pipeline (Fig \ref{fig:pipeline-phase1}, shared legend in Fig \ref{fig:pipeline-key}) consists of steps which can be applied independently at the level of sequencing runs.
For large sequencing runs (\textgreater{} 96 samples) the pipeline processes samples within each sequencing run in batches of at most 96 samples for steps which do not require information from other samples.
The steps in phase 1 are as follows:

\emph{Orientation, primer detection and quality filtering.}
The OptimOTU pipeline can process reads with a variety of orientations, depending on the indexing scheme used in the sequencing run and specified in the configuration file with the option \texttt{orient}.
Possible values are ``fwd'', ``rev'', ``mixed'', and ``custom''.
In the ``fwd'' orientation, the forward primer is expected at the 5' end of R1 and the reverse primer at the 5' end of R2.
In the ``rev'' orientation, the reverse primer is expected at the 5' end of R1 and the forward primer at the 5' end of R2, and the ASV sequence is reverse-complemented after denoising and merging of each pair.
In the ``mixed'' orientation, both orientations are expected and searched for in every sample.
Reads detected in the ``fwd'' and ``rev'' orientation are processed separately through phase 1 of the pipeline until pair merging, at which point the reverse-oriented reads are reverse-complemented.
In the ``custom'' orientation, the user must supply a custom sample table (see above) which includes a column ``orient'' with values ``fwd'', ``rev'', or ``mixed'' for each sample.
This is primarily used for certain indexing schemes where different samples are sequenced in different orientations to increase multiplexing capacity and sequence diversity across the flow cell.

Primers are specified in the configuration file with the options \texttt{forward\_primer} and \texttt{reverse\_primer}.
The expected primer at the 5' end of each read is searched for using Cutadapt (Martin 2011).
Read pairs where the expected primer is not found are discarded.
The reverse-complemented sequence of the other primer is also searched for at the 3' end of each read, which may occur in marker sequences with highly variable length such as ITS.
If a primer sequence is found it is removed if found, but its presence is not required.
Settings for Cutadapt are specified in the configuration with sub-options to \texttt{trimming}.
In particular, the action to take on the primer sequence is specified with the option \texttt{action}.
Although Cutadapt has several options for this, the OptimOTU pipeline supports two, ``trim'' and ``retain''.
Option ``trim'' removes the primer sequence from the read immediately, while ``retain'' leaves it in place, but trims any bases before the primer sequence.
If option ``retain'' is used, the trimming still occurs, but it is applied to the master ASV list at the beginning of phase 2.
This means that the primer sequence is present during the denoising stage.
The conserved endpoints may aid in pairwise alignment of highly variable sequences during denoising.
However, this option is not recommended in most situations, and in particular is not suitable when primers with degenerate bases are used.

At the same time as primer searching, reads are trimmed and quality filtered.
Options are user configurable, but a typical set of options trims low quality bases from the 5' and 3' ends of each read, removes reads with any ambiguous bases after trimming, and discard reads which are less than a specified length.
Options controlling primer matching, trimming, and filtering can optionally be included in the custom sample table.
This option is useful when different sequencing runs have been sequenced with different settings or instruments.

\emph{Denoising:}
The denoising steps are performed using the DADA2 package (Callahan et al. 2016).
Trimmed read pairs are first subjected to an additional round of quality filtering using DADA2's \texttt{fastqPairedFilter()}, based on the expected number of errors in the read.
Error profiles are then learned separately for each sequencing run, read, and orientation using the \texttt{learnErrors()} function.
Sequences with binned quality scores, as produced by newer Illumina sequencers, are automatically detected, and the error model is adjusted accordingly.
Denoising is then performed using the \texttt{dada()} function, and read pairs are merged using the \texttt{mergePairs()} function.
The denoised R1 and R2 reads from any reverse-oriented read pairs are swapped at this stage.
Thus the merged, denoised ASVs are all in the forward orientation.

\emph{ASV table construction:}
The DADA2 standard storage format for the ASV × sample table is as an R matrix of integer read counts with sequences as column names.
In projects with a very large number of samples and high diversity of ASVs, this structure has prohibitively high memory requirements.
In the OptimOTU pipeine, the unique ASV sequences are stored in a single gzipped FASTA file, which is indexed using the open-source tool FastqIndEx (\url{https://github.com/DKFZ-ODCF/FastqIndEx}) to allows fast random access to sequences.
The matrix of ASV read counts in each sample is then constructed and stored in coordinate list (COO) sparse matrix format, as a three column R data frame where the columns represent a sample identifier, index of the sequence in the master ASV file, and number of reads.
Merging ASV\textsubscript{×}sample tables for multiple samples and sequencing runs is then a simple matter of row-concatenating the tables for the individual samples.

\emph{Tag-switching detection:}
Before assembly of the final global sample table, potential cases of tag-switching are optionally detected and removed using an R implementation of the UNCROSS2 algorithm (Edgar 2018) which operates on the COO matrix format.
Because tag-switching is unlikely between samples which were not sequenced together, UNCROSS2 is applied separately for each sequencing run.

\emph{De novo chimera detection:}
Two different chimera detection methods are applied in the pipeline.
The first is an adaptation of the consensus algorithm implemented in DADA2's \texttt{isBimeraDenovoTable()} function.
This function first runs \emph{de novo} chimera detection individually on the ASVs from each sample, and then combines the results into a master list of potential bimeras.
The potential bimeras are then flagged for removal if they were detected as chimeras in a majority (DADA2 default 90\% - 1) of the samples in which they occur.
The use of a consensus between samples is intended to reduce the number of false positives and preserve more real ASV diversity in the final table.
The OptimOTU pipeline uses a partial re-implementation of this method, where the DADA2 intenal function \texttt{C\_table\_bimera2} is used to flag potential bimeras in small subsets of sequences, but then the results are combined across all samples before determining the set of consensus chimeras.
Because the \emph{de novo} chimera detection is performed within samples, it is nominally part of phase 1 of the pipeline, even though it is computationally downstream of the candidate ASV table construction.

\begin{figure}
\centering
\includegraphics{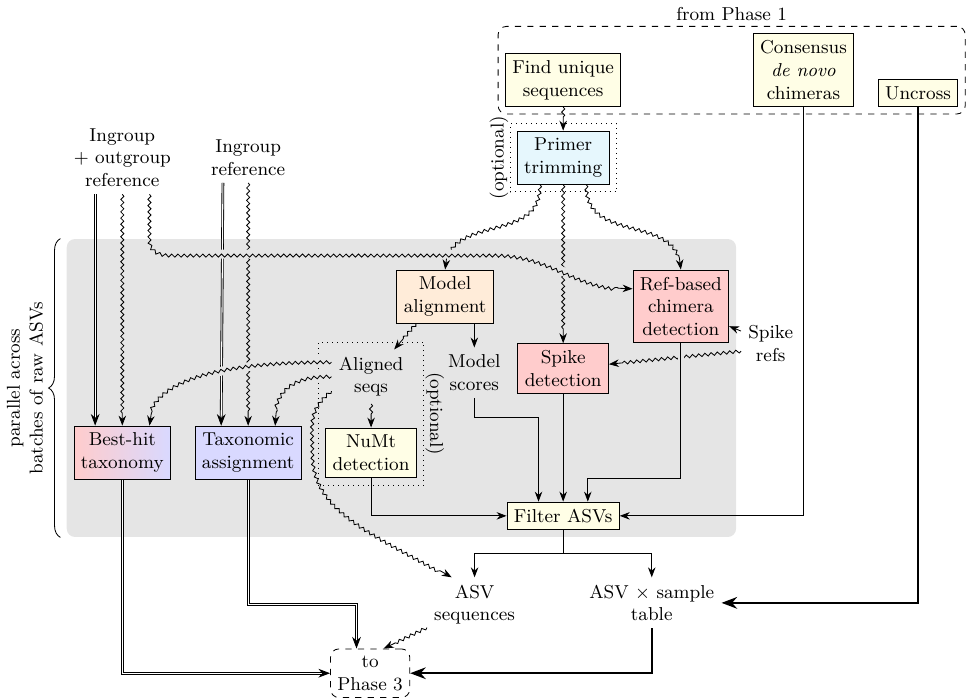}
\caption{\label{fig:pipeline-phase2}Phase 2 of the OptimOTU pipeline.}
\end{figure}

\subsubsection{Phase 2}\label{phase-2}

Phase 2 (Fig \ref{fig:pipeline-phase2}, shared legend in Fig \ref{fig:pipeline-key}) consists of steps which are performed on batches from the master list of candidate ASVs.
The ordering of the master list, as well as the batch structure, are cached between runs, so that if the pipeline is re-run with additional samples, ASVs encountered in previous runs are not re-processed.
Many of the steps in phase 2 are intended to remove non-target sequences such as spike-ins, chimeras, and paralogs.
Taxonomic identification is also performed at this stage.
Although it would in principle be possible to perform these operations serially for each batch, so that sequences discarded in earlier steps do not need to be processed in later steps, the pipeline instead performs all operations on all sequences.
This allows results to be compared for different artifact detection steps.
The steps in phase 2 are as follows:

\emph{Spike detection:}
The pipeline includes an optional step to detect spike-ins, which are DNA standards added to the samples at a stage prior to PCR amplification.
They may be used to estimate the quantity of sample DNA (Ovaskainen et al. 2020) or as a quality control measure.
A FASTA file containing spike sequences may be specified by the user, and spike detection is performed on the candidate ASV list using the ``usearch\_global'' command (Edgar 2010) in VSEARCH (Rognes et al. 2016).
The detected spike sequences are removed from the final ASV table, but the number of reads attributed to spike sequences in each sample is reported in the final read count table (see ``Outputs'', below) for downstream analysis.
The supplied spike sequences are also added to the reference database for chimera detection (below).

\emph{Reference-based chimera detection:}
The second chimera detection method is the original reference-based UCHIME algorithm (Edgar et al. 2011), as implemented in VSEARCH (Rognes et al. 2016).
The reference database used for this step is user-configurable, and if a set of spike sequences has been specified then they are also included.

\emph{Model alignment:}
Candidate ASV sequences can also be aligned to either a profile HMM (Krogh et al. 1994) or CM (Eddy and Durbin 1994) of the target marker gene.
HMMs may be used for any marker gene, while CMs are specific to non-coding RNA genes with a conserved secondary structure, such as ribosomal RNA genes.
Model alignment can be used in three different ways by the pipeline.
First, the model can be used to detect non-target or partial sequences using the model alignment score for each ASV sequence.
In this use, the alignment itself is not stored, but low-scoring sequences can be flagged as likely artifacts.
Even amplicons of highly variable marker genes such as ITS can be aligned to a model for filtering, although in this case the alignment score is mostly derived from the flanking conserved regions.
Second, for protein coding genes, the distribution of gaps in the model alignment can be used to detect frame shifts in the ASV sequences.
Such frame shifts are unlikely to occur in real protein coding sequences, and can be used to detect nuclear mitochondrial (NuMt; see below) or other pseudogenes (Porter and Hajibabaei 2021).
Alternatively, apparent frame shift mutations may be due to PCR or sequencing errors, which should also be removed.
Finally, the model alignment may be used as the input to the taxonomic alignment and classification steps, allowing the use of fast Hamming distance calculations.
Model alignment is performed using the nhmmer command from the HMMER package (Eddy 2011) for HMMs or the cmalign command from the Infernal package (Nawrocki 2014) for CMs.

\emph{NuMt detection:}
Nuclear mitochondrial pseudogenes (NuMts) are sequences which are derived from the mitochondrial genome but have been inserted into the nuclear genome, where they are inactive.
They are common in many eukaryotic genomes, and can be difficult to distinguish from true mitochondrial sequences, which may inflate OTU counts based on mitochondrial marker genes such as COI (Song et al. 2008).
The OptimOTU pipeline includes an optional pseudogene filter for use especially for mitochondrial protein-coding genes such as COI, which uses two methods to detect pseudogenes.
The first, alluded to above, is the detection of frame shifts in the model alignment.
These are defined as insertions or deletion which are not a multiple of three.
Full codon insertions and deletions relative to the HMM are not considered as pseudogenes, since these are plausibly biological in origin, and are known to occur in some taxa (Pentinsaari et al. 2016).
The second method is to search for sequences which include in-frame stop codons, which are also extremely unlikely to occur in real protein sequences of conserved genes.
By default the OptimOTU pipeline uses the invertebrate mitochondrial stop codons TAA and TAG.

\emph{Taxonomic identification:}
The OptimOTU pipeline uses two implementations of Protax (Somervuo et al. 2016) for taxonomic assignment:
Protax-Fungi (Abarenkov et al. 2018) for fungal ITS sequences, and Protax-Animal (Roslin et al. 2022; R. Li et al. 2024) for metazoan COI sequences.
It is also possible for the user to specify a custom Protax model for other marker genes or taxonomic groups.
The Protax-Fungi implementation is more appropriate for genes which are not easily aligned across the entire taxonomic group of interest, while the Protax-Animal implementation is more appropriate for genes which can be reliably aligned across all query sequences using an HMM or CM.
Protax produces a probability distribution across different possible taxa for each ASV at each rank, while the OptimOTU clustering algorithm requires a single taxonomic assignment (or ``unknown'') for each ASV.
The pipeline uses two alternate probability thresholds, 50\% and 90\%, which are designated ``probable'' and ``reliable'' after Somervuo et al. (2017).
All analyses downstream of taxonomic identification, most notably OptimOTU clustering, are performed separately for each of these two probability thresholds, and both results are reported in the final outputs.

\emph{Outgroup identification:}
Both Protax implementations use reference databases which are restricted to the target taxonomic kingdoms.
However, sequences from other kingdoms, referred to as ``outgroup'' sequences, are often present in environmentally derived datasets.
These sequences are identified using search against more inclusive reference databases (Unite all-eukaryotes release for ITS; BOLD for COI).
Search for unaligned sequences is performed using the VSEARCH implementation of \texttt{usearch\_global} (Edgar 2010; Rognes et al. 2016), a heuristic method, while search for aligned sequences is performed using the fast Hamming distance calculations of the ProtaxAnimal implementation.
ASVs are classified at this stage as ``ingroup'' sequences if their closest match in the expanded reference is to the target taxonomic group, as ``outgroup'' sequences if their closest match is to a different taxonomic group, and as ``unknown'' if their closest match is to a sequence which is itself not identified, or if there is no match at the specified search threshold (default genetic distance 0.2).
Outgroup and unknown sequences are not removed until after the clustering stage (see ``Outgroup removal'' below).

\emph{Final ASV table:}
The final ASV table is constructed by removing sequences flagged by the spike-in detection, chimera detection, model score filtering, and NuMt detection steps.
Remaining ASVs are sorted from highest to lowest according to prevalence (number of samples they occur in), with ties broken by total abundance (number of reads across all samples), then variance in abundance across samples.
Remaining ties are broken by sorting the hash of the sequences, which is pseudorandom but consistent across runs.
The ASVs are then given identifiers according to their position in the sorted list, such that low numbers are given to ASVs which are more prevalent and abundant across samples, and which show stronger ecological signal in the data.

\begin{figure}
\centering
\includegraphics{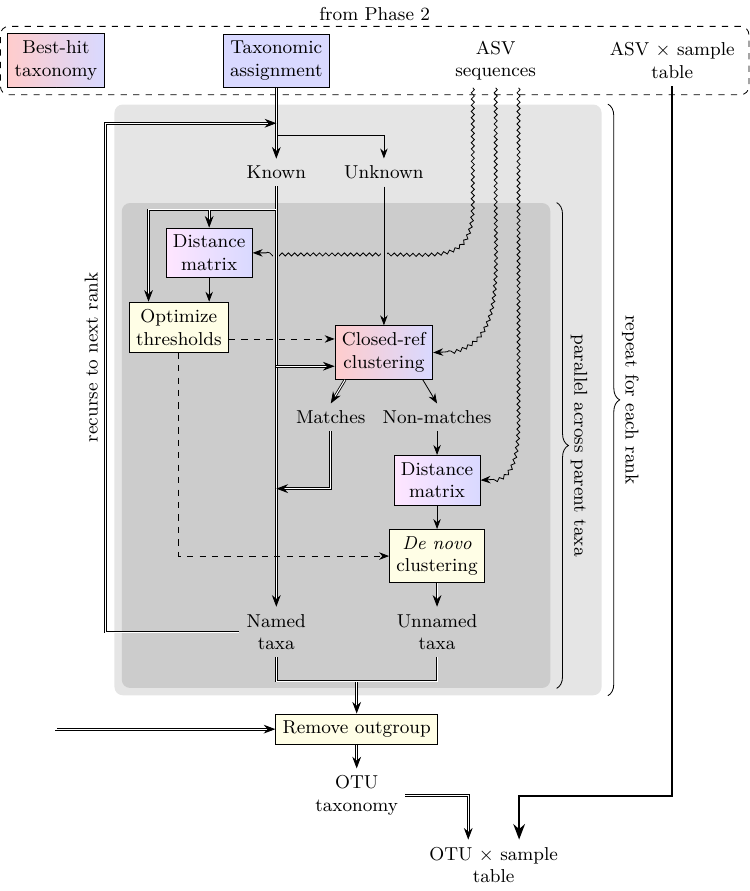}
\caption{\label{fig:pipeline-phase3}Phase 3 of the OptimOTU pipeline.}
\end{figure}

\subsubsection{Phase 3}\label{phase-3}

Phase 3 (Fig \ref{fig:pipeline-phase3}, shared legend in Fig \ref{fig:pipeline-key}) consists of steps which are performed on the final ASV list, using the taxonomy and outgroup information produced in phase 2.
The steps in phase 3 are as follows:

\emph{Taxonomically guided clustering:}
Clustering of the final ASV list is performed using the OptimOTU clustering algorithm, as described above.
The procedure is not performed using a single call to the top-level \texttt{optimotu()} function, but instead the closed reference and \emph{de novo} clustering steps are defined as separate targets in the pipeline for each rank and parent taxon, allowing the work to be split across multiple nodes.
For aligned sequences, the pairwise distances are calculated using the Hamming distance, while for unaligned sequences pairwise distances are calculated by USEARCH.

\emph{Outgroup removal:}
After clustering, pseudotaxa at the highest taxonomic rank (by default phylum) are assessed for the presence of outgroup sequences.
Pseudotaxa which contain more outgroup and unknown sequences than ingroup sequences are removed from the final OTU table.

\subsubsection{Outputs}\label{outputs}

The final outputs of the pipeline are a set of tables, each written in both TSV format for access by external tools, and RDS format for easy loading in R, as well as sequences in gzipped FASTA format.
The names of the individual files are always the same, enabling easy access to the results from standardized analysis scripts.
However, to ensure identifiability and prevent loss of data, at the end of each run, all of the output files are zipped into a single archive, which is named according to the date of the run and the name of the project as specified in the configuration file.
The content and formatting of each file is described below.

\emph{ASV table:} Two files named \texttt{asv\_table.rds} and \texttt{asv\_table.tsv}.
The ASV table is a sparse matrix with five columns: \texttt{sample} (string), \texttt{seqrun} (string), \texttt{seq\_id} (string), \texttt{seq\_idx} (integer), and \texttt{nread} (integer).
The \texttt{sample} and \texttt{seqrun} columns together identify a fastq file pair in which sequences mapping to a particular ASV were found.
The \texttt{seq\_id} column is the sequence identifier, which takes the form ``ASVXXXX'', where XXXX is a zero-padded integer, and the number of digits is adjusted to fit all ASVs in the dataset.
The \texttt{seq\_idx} column is the index of the sequence in the master ASV file.
Its value is typically the same as the numerical part of \texttt{seq\_id}.
Finally, the \texttt{nread} column is the number of reads in the sample which map to the ASV.

\emph{ASV taxonomy:} Four files named \texttt{asv\_taxonomy\_\{confidence\}.\{ext\}}, where \texttt{\{confidence\}} is either \texttt{plausible} or \texttt{reliable}, and \texttt{\{ext\}} is either \texttt{rds} or \texttt{tsv}.
The ASV taxonomy tables contains a variable number of columns.
The first column, \texttt{seq\_id}, is the same as in the ASV table, but only ASVs which were retained in the outgroup removal step are included.
In the RDS formatted files, the sequence identifiers are stored as row names instead of a data column.
The rest of of the columns are named for the taxonomic ranks, and contain the taxonomic assignments.
Both named taxa (from taxonomic identifications) and pseudotaxa (from \emph{de novo} clustering) are included.
The \texttt{plausible} and \texttt{reliable} versions of the table are based on taxonomic assignments at the 50\% and 90\% probability thresholds, respectively.

\emph{OTU taxonomy:} Four files named \texttt{otu\_taxonomy\_\{confidence\}.\{ext\}}, where \texttt{\{confidence\}} is either \texttt{plausible} or \texttt{reliable}, and \texttt{\{ext\}} is either \texttt{rds} or \texttt{tsv}.
The OTU taxonomy tables are similar to the ASV taxonomy tables, but all ASVs in each OTU (i.e., those with identical taxonomy) are combined into a single row.
The \texttt{seq\_id} column is the OTU identifier, which is of the form ``OTUXXXX'', where XXXX is a zero-padded integer.
Like the ASV taxonomy tables, the RDS formatted files store the sequence identifiers as row names.
There is also a \texttt{ref\_seq\_id} column, which is the ASV identifier of the representative sequence for the OTU.
Columns \texttt{nsample} and \texttt{nread} contain the total number of samples and reads in which the OTU was found.
The rest of the columns are named for the taxonomic ranks, and contain the taxonomic assignments.

\emph{OTU table:} Four files named \texttt{otu\_table\_sparse\_\{confidence\}.\{ext\}}, where \texttt{\{confidence\}} is either \texttt{plausible} or \texttt{reliable}, and \texttt{\{ext\}} is either \texttt{rds} or \texttt{tsv};
optionally also four files named \texttt{otu\_table\_\{confidence\}.\{ext\}}.
The default, ``sparse'' version of the OTU table has six columns.
The first four columns, \texttt{sample}, \texttt{seqrun}, \texttt{seq\_id}, and \texttt{nread}, are the same as the columns of those names in the ASV table, except that the \texttt{seq\_id} column is the OTU identifier instead of the ASV identifier.
The last two columns are transformed abundance measurements.
Column \texttt{fread} is the fraction of all reads in the sample which belong to the OTU in question.
Column \texttt{w} is the spike-weighted read abundance, defined as the number of reads belonging to the OTU divided by one plus the number of reads in the sample which were identified as spike sequences.
The ``dense'' version of the OTU table is an integer matrix with rows corresponding to samples and columns corresponding to OTU identifiers.
If the sample names are not all unique, then the row names include both the sample and the sequencing run.
The matrix values are then the number of reads for each OTU in each sample.
This matrix is not generated by default because in large projects it may be inconveniently large, but it can be activated by including the option \texttt{dense\_table:\ yes} in the configuration file.

\emph{OTU reference sequences:} Two files named \texttt{otu\_plausible.fasta.gz} and \texttt{otu\_reliable.fasta.gz}.
These files contain the representative sequences for each OTU, in FASTA format.
The sequences are named according to the OTU identifier, and the sequence itself is the sequence of the ASV with the smallest index.
Because of the ASV sorting procedure, this is the most prevalent and abundant ASV in the OTU.

\emph{Read counts:} Four files named \texttt{read\_counts\_\{confidence\}.\{ext\}}, where \texttt{\{confidence\}} is either \texttt{plausible} or \texttt{reliable}, and \texttt{\{ext\}} is either \texttt{rds} or \texttt{tsv}.
The read counts table contains the number of reads in each sample which were present after each stage of the pipeline.
Samples are identified with the same \texttt{sample} and \texttt{seqrun} columns as in the ASV and OTU tables.
The remainder of the columns are named \texttt{\{stage\}\_nread}, where \texttt{\{stage\}} is the name of the stage in the pipeline.
Stages which are always included are ``raw'' for the input files, ``trim'' for sequences which passed the Cutadapt orientation and filtering stage, ``filt'' for sequences which passed the expected-errors filter, ``denoise'', ``nochim1'' for sequences not flagged as chimeric by the consensus \emph{de novo} method, ``nochim2'' for sequences not flagged as chimeric by the reference-based method, and ``ingroup'' for the final set of OTU sequences which were classified as ingroup by the outgroup detection step.
If the tag-switching detection step is included, then the ``uncross'' stage is also included.
If the spike-in detection step is included, then the ``spike'' and ``nospike'' stages are also included.
If the model alignment step is included, then the ``full\_length'' stage is also included.

\section*{Software availability}\label{software-availability}
\addcontentsline{toc}{section}{Software availability}

The R packages \texttt{optimotu} and \texttt{optimotu.pipeline} are available under the MIT license on GitHub at \url{https://github.com/brendanf/optimotu} and \url{https://github.com/brendanf/optimotu.pipeline}, respectively.
The \texttt{targets} pipeline for the OptimOTU pipeline is available at \url{https://github.com/brendanf/optimotu_targets}.

\section*{Funding}\label{funding}
\addcontentsline{toc}{section}{Funding}

This work was supported by the Research Council of Finland (grant no. 336212 and 345110); and the European Union: the European Research Council (ERC) under the European Union's Horizon 2020 research and innovation programme (grant agreement No 856506; ERC-synergy project LIFEPLAN) and the HORIZON-CL6-2021-BIODIV-01 project 101059492 (Biodiversity Genomics Europe).

\section*{Acknowledgements}\label{acknowledgements}
\addcontentsline{toc}{section}{Acknowledgements}

The authors would like to thank Mira Kajanus, Domenica Naranjo-Orrico, and Skylar Burg for pipeline testing. Bess Hardwick and Deirdre Kerdraon provided project management for the GSSP and LIFEPLAN projects under which OptimOTU was developed. Computational resources were provided by CSC -- IT Center For Science,

\section*{References}\label{references}
\addcontentsline{toc}{section}{References}

\phantomsection\label{refs}


\begin{CSLReferences}{1}{0}
\bibitem[\citeproctext]{ref-abarenkov2010}
Abarenkov, Kessy, R. Henrik Nilsson, Karl-Henrik Larsson, Ian J. Alexander, Ursula Eberhardt, Susanne Erland, Klaus Høiland, et al. 2010. {``The {UNITE} Database for Molecular Identification of Fungi -- Recent Updates and Future Perspectives.''} \emph{New Phytologist} 186 (2): 281--85. \url{https://doi.org/10.1111/j.1469-8137.2009.03160.x}.

\bibitem[\citeproctext]{ref-abarenkov2018a}
Abarenkov, Kessy, Panu Somervuo, R. Henrik Nilsson, Paul M. Kirk, Tea Huotari, Nerea Abrego, and Otso Ovaskainen. 2018. {``Protax-Fungi: A Web-Based Tool for Probabilistic Taxonomic Placement of Fungal Internal Transcribed Spacer Sequences.''} \emph{New Phytologist} 220 (2): 517--25. \url{https://doi.org/10.1111/nph.15301}.

\bibitem[\citeproctext]{ref-amir2017}
Amir, Amnon, Daniel McDonald, Jose A. Navas-Molina, Evguenia Kopylova, James T. Morton, Zhenjiang Zech Xu, Eric P. Kightley, et al. 2017. {``Deblur {Rapidly Resolves Single-Nucleotide Community Sequence Patterns}.''} \emph{mSystems} 2 (2): e00191--16. \url{https://doi.org/10.1128/mSystems.00191-16}.

\bibitem[\citeproctext]{ref-anslan2017}
Anslan, Sten, Mohammad Bahram, Indrek Hiiesalu, and Leho Tedersoo. 2017. {``{PipeCraft}: {Flexible} Open-Source Toolkit for Bioinformatics Analysis of Custom High-Throughput Amplicon Sequencing Data.''} \emph{Molecular Ecology Resources} 17 (6): e234--40. \url{https://doi.org/10.1111/1755-0998.12692}.

\bibitem[\citeproctext]{ref-bokulich2018}
Bokulich, Nicholas A., Benjamin D. Kaehler, Jai Ram Rideout, Matthew Dillon, Evan Bolyen, Rob Knight, Gavin A. Huttley, and J. Gregory Caporaso. 2018. {``Optimizing Taxonomic Classification of Marker-Gene Amplicon Sequences with {QIIME} 2's Q2-Feature-Classifier Plugin.''} \emph{Microbiome} 6 (1): 90. \url{https://doi.org/10.1186/s40168-018-0470-z}.

\bibitem[\citeproctext]{ref-callahan2017}
Callahan, Benjamin J., Paul J. McMurdie, and Susan P. Holmes. 2017. {``Exact Sequence Variants Should Replace Operational Taxonomic Units in Marker-Gene Data Analysis.''} \emph{The ISME Journal} 11 (12): 2639--43. \url{https://doi.org/10.1038/ismej.2017.119}.

\bibitem[\citeproctext]{ref-callahan2016}
Callahan, Benjamin J., Paul J. McMurdie, Michael J. Rosen, Andrew W. Han, Amy Jo A. Johnson, and Susan P. Holmes. 2016. {``{DADA2}: {High-resolution} Sample Inference from {Illumina} Amplicon Data.''} \emph{Nature Methods} 13 (7): 581--83. \url{https://doi.org/10.1038/nmeth.3869}.

\bibitem[\citeproctext]{ref-ching2024}
Ching, Travers. 2024. {``Qs: {Quick} Serialization of {R} Objects.''} \url{https://github.com/qsbase/qs}.

\bibitem[\citeproctext]{ref-conda_contributors_conda_A_system-level}
conda contributors. n.d. {``Conda: {A} System-Level, Binary Package and Environment Manager Running on All Major Operating Systems and Platforms.''} \url{https://github.com/conda/conda}.

\bibitem[\citeproctext]{ref-czech2020}
Czech, Lucas, Pierre Barbera, and Alexandros Stamatakis. 2020. {``Genesis and {Gappa}: Processing, Analyzing and Visualizing Phylogenetic (Placement) Data.''} \emph{Bioinformatics} 36 (10): 3263--65. \url{https://doi.org/10.1093/bioinformatics/btaa070}.

\bibitem[\citeproctext]{ref-dondoshansky2000blastclust}
Dondoshansky, I, and Y Wolf. 2000. {``{BLASTCLUST} - {BLAST} Score-Based Single Linkage Clustering.''} National Center for Biotechnology Information. \url{ftp://ftp.ncbi.nih.gov/blast/documents/blastclust.html}.

\bibitem[\citeproctext]{ref-eddy2011}
Eddy, Sean R. 2011. {``Accelerated {Profile HMM Searches}.''} \emph{PLOS Computational Biology} 7 (10): e1002195. \url{https://doi.org/10.1371/journal.pcbi.1002195}.

\bibitem[\citeproctext]{ref-eddy1994}
Eddy, Sean R., and Richard Durbin. 1994. {``{RNA} Sequence Analysis Using Covariance Models.''} \emph{Nucleic Acids Research} 22 (11): 2079--88. \url{https://doi.org/10.1093/nar/22.11.2079}.

\bibitem[\citeproctext]{ref-edgar2010}
Edgar, Robert C. 2010. {``Search and Clustering Orders of Magnitude Faster Than {BLAST}.''} \emph{Bioinformatics} 26 (19): 2460--61. \url{https://doi.org/10.1093/bioinformatics/btq461}.

\bibitem[\citeproctext]{ref-edgar2016a}
---------. 2016a. {``{SINTAX}: A Simple Non-{Bayesian} Taxonomy Classifier for {16S} and {ITS} Sequences.''} \emph{bioRxiv}, September, 074161. \url{https://doi.org/10.1101/074161}.

\bibitem[\citeproctext]{ref-edgar2016}
---------. 2016b. {``{UNOISE2}: Improved Error-Correction for {Illumina 16S} and {ITS} Amplicon Sequencing.''} \emph{bioRxiv}, October, 081257. \url{https://doi.org/10.1101/081257}.

\bibitem[\citeproctext]{ref-edgar2018c}
---------. 2018. {``{UNCROSS2}: Identification of Cross-Talk in {16S rRNA OTU} Tables.''} August 27, 2018. \url{https://doi.org/10.1101/400762}.

\bibitem[\citeproctext]{ref-edgar2011}
Edgar, Robert C., Brian J. Haas, Jose C. Clemente, Christopher Quince, and Rob Knight. 2011. {``{UCHIME} Improves Sensitivity and Speed of Chimera Detection.''} \emph{Bioinformatics} 27 (16): 2194--2200. \url{https://doi.org/10.1093/bioinformatics/btr381}.

\bibitem[\citeproctext]{ref-fowlkes1983}
Fowlkes, E. B., and C. L. Mallows. 1983. {``A {Method} for {Comparing Two Hierarchical Clusterings}.''} \emph{Journal of the American Statistical Association} 78 (383): 553--69. \url{https://doi.org/10.1080/01621459.1983.10478008}.

\bibitem[\citeproctext]{ref-guillou2013}
Guillou, Laure, Dipankar Bachar, Stéphane Audic, David Bass, Cédric Berney, Lucie Bittner, Christophe Boutte, et al. 2013. {``The {Protist Ribosomal Reference} Database ({PR2}): A Catalog of Unicellular Eukaryote Small Sub-Unit {rRNA} Sequences with Curated Taxonomy.''} \emph{Nucleic Acids Research} 41 (January): D597--604. \url{https://doi.org/10.1093/nar/gks1160}.

\bibitem[\citeproctext]{ref-hubert1985}
Hubert, Lawrence, and Phipps Arabie. 1985. {``Comparing Partitions.''} \emph{Journal of Classification} 2 (1): 193--218. \url{https://doi.org/10.1007/BF01908075}.

\bibitem[\citeproctext]{ref-jamy2020}
Jamy, Mahwash, Rachel Foster, Pierre Barbera, Lucas Czech, Alexey Kozlov, Alexandros Stamatakis, Gary Bending, Sally Hilton, David Bass, and Fabien Burki. 2020. {``Long-Read Metabarcoding of the Eukaryotic {rDNA} Operon to Phylogenetically and Taxonomically Resolve Environmental Diversity.''} \emph{Molecular Ecology Resources} 20 (2): 429--43. \url{https://doi.org/10.1111/1755-0998.13117}.

\bibitem[\citeproctext]{ref-kauserud2023}
Kauserud, Håvard. 2023. {``{ITS} Alchemy: {On} the Use of {ITS} as a {DNA} Marker in Fungal Ecology.''} \emph{Fungal Ecology}, July, 101274. \url{https://doi.org/10.1016/j.funeco.2023.101274}.

\bibitem[\citeproctext]{ref-klik2022}
Klik, Mark. 2022. {``Fst: {Lightning} Fast Serialization of Data Frames.''} \url{http://www.fstpackage.org}.

\bibitem[\citeproctext]{ref-krogh1994}
Krogh, Anders, Michael Brown, I. Saira Mian, Kimmen Sjölander, and David Haussler. 1994. {``Hidden {Markov Models} in {Computational Biology}: {Applications} to {Protein Modeling}.''} \emph{Journal of Molecular Biology} 235 (5): 1501--31. \url{https://doi.org/10.1006/jmbi.1994.1104}.

\bibitem[\citeproctext]{ref-landau2021}
Landau, William Michael. 2021. {``The Targets {R} Package: A Dynamic {Make-like} Function-Oriented Pipeline Toolkit for Reproducibility and High-Performance Computing.''} \emph{Journal of Open Source Software} 6 (57): 2959. \url{https://doi.org/10.21105/joss.02959}.

\bibitem[\citeproctext]{ref-landau2024}
Landau, William Michael, Michael Gilbert Levin, and Brendan Furneaux. 2024. {``Crew.cluster: {Crew} Launcher Plugins for Traditional High-Performance Computing Clusters.''} \url{https://wlandau.github.io/crew.cluster/}.

\bibitem[\citeproctext]{ref-lanzen2012}
Lanzén, Anders, Steffen L. Jørgensen, Daniel H. Huson, Markus Gorfer, Svenn Helge Grindhaug, Inge Jonassen, Lise Øvreås, and Tim Urich. 2012. {``{CREST} -- {Classification Resources} for {Environmental Sequence Tags}.''} \emph{PLOS ONE} 7 (11): e49334. \url{https://doi.org/10.1371/journal.pone.0049334}.

\bibitem[\citeproctext]{ref-li2024}
Li, Roy, Sujeevan Ratnasingham, Iuliia Zarubiieva, Panu Somervuo, and Graham W. Taylor. 2024. {``{PROTAX-GPU}: A Scalable Probabilistic Taxonomic Classification System for {DNA} Barcodes.''} \emph{Philosophical Transactions of the Royal Society B: Biological Sciences} 379 (1904): 20230124. \url{https://doi.org/10.1098/rstb.2023.0124}.

\bibitem[\citeproctext]{ref-li2006}
Li, Weizhong, and Adam Godzik. 2006. {``Cd-Hit: A Fast Program for Clustering and Comparing Large Sets of Protein or Nucleotide Sequences.''} \emph{Bioinformatics (Oxford, England)} 22 (13): 1658--59. \url{https://doi.org/10.1093/bioinformatics/btl158}.

\bibitem[\citeproctext]{ref-mahe2015}
Mahé, Frédéric, Torbjørn Rognes, Christopher Quince, Colomban de Vargas, and Micah Dunthorn. 2015. {``Swarm V2: Highly-Scalable and High-Resolution Amplicon Clustering.''} \emph{PeerJ} 3: e1420. \url{https://doi.org/10.7717/peerj.1420}.

\bibitem[\citeproctext]{ref-marco-sola2021}
Marco-Sola, Santiago, Juan Carlos Moure, Miquel Moreto, and Antonio Espinosa. 2021. {``Fast Gap-Affine Pairwise Alignment Using the Wavefront Algorithm.''} \emph{Bioinformatics} 37 (4): 456--63. \url{https://doi.org/10.1093/bioinformatics/btaa777}.

\bibitem[\citeproctext]{ref-martin2011}
Martin, Marcel. 2011. {``Cutadapt Removes Adapter Sequences from High-Throughput Sequencing Reads.''} \emph{EMBnet.journal} 17 (1): 10--12. \url{https://doi.org/10.14806/ej.17.1.200}.

\bibitem[\citeproctext]{ref-matthews1975}
Matthews, B. W. 1975. {``Comparison of the Predicted and Observed Secondary Structure of {T4} Phage Lysozyme.''} \emph{Biochimica Et Biophysica Acta (BBA) - Protein Structure} 405 (2): 442--51. \url{https://doi.org/10.1016/0005-2795(75)90109-9}.

\bibitem[\citeproctext]{ref-murali2018}
Murali, Adithya, Aniruddha Bhargava, and Erik S. Wright. 2018. {``{IDTAXA}: A Novel Approach for Accurate Taxonomic Classification of Microbiome Sequences.''} \emph{Microbiome} 6 (1): 140. \url{https://doi.org/10.1186/s40168-018-0521-5}.

\bibitem[\citeproctext]{ref-nawrocki2014}
Nawrocki, Eric P. 2014. {``Annotating {Functional RNAs} in {Genomes Using Infernal}.''} In \emph{{RNA Sequence}, {Structure}, and {Function}: {Computational} and {Bioinformatic Methods}}, edited by Jan Gorodkin and Walter L. Ruzzo, 163--97. Methods in {Molecular Biology}. Totowa, NJ: Humana Press. \url{https://doi.org/10.1007/978-1-62703-709-9_9}.

\bibitem[\citeproctext]{ref-nilsson2008}
Nilsson, R. Henrik, Erik Kristiansson, Martin Ryberg, Nils Hallenberg, and Karl-Henrik Larsson. 2008. {``Intraspecific {ITS Variability} in the {Kingdom Fungi} as {Expressed} in the {International Sequence Databases} and {Its Implications} for {Molecular Species Identification}.''} \emph{Evolutionary Bioinformatics} 4 (January): EBO.S653. \url{https://doi.org/10.4137/EBO.S653}.

\bibitem[\citeproctext]{ref-nolet2023}
Nolet, Corey J., Divye Gala, Alex Fender, Mahesh Doijade, Joe Eaton, Edward Raff, John Zedlewski, Brad Rees, and Tim Oates. 2023. {``{cuSLINK}: {Single-Linkage Agglomerative Clustering} on~the~{GPU}.''} In \emph{Machine {Learning} and {Knowledge Discovery} in {Databases}: {Research Track}}, edited by Danai Koutra, Claudia Plant, Manuel Gomez Rodriguez, Elena Baralis, and Francesco Bonchi, 711--26. Lecture {Notes} in {Computer Science}. Cham: Springer Nature Switzerland. \url{https://doi.org/10.1007/978-3-031-43412-9_42}.

\bibitem[\citeproctext]{ref-ovaskainen2024}
Ovaskainen, Otso, Nerea Abrego, Brendan Furneaux, Bess Hardwick, Panu Somervuo, Isabella Palorinne, Nigel R. Andrew, et al. 2024. {``Global {Spore Sampling Project}: {A} Global, Standardized Dataset of Airborne Fungal {DNA}.''} \emph{Scientific Data} 11 (1): 561. \url{https://doi.org/10.1038/s41597-024-03410-0}.

\bibitem[\citeproctext]{ref-ovaskainen2020}
Ovaskainen, Otso, Nerea Abrego, Panu Somervuo, Isabella Palorinne, Bess Hardwick, Juha-Matti Pitkänen, Nigel R. Andrew, et al. 2020. {``Monitoring {Fungal Communities With} the {Global Spore Sampling Project}.''} \emph{Frontiers in Ecology and Evolution} 7. \url{https://doi.org/10.3389/fevo.2019.00511}.

\bibitem[\citeproctext]{ref-pentinsaari2016}
Pentinsaari, Mikko, Heli Salmela, Marko Mutanen, and Tomas Roslin. 2016. {``Molecular Evolution of a Widely-Adopted Taxonomic Marker ({COI}) Across the Animal Tree of Life.''} \emph{Scientific Reports} 6 (1, 1): 35275. \url{https://doi.org/10.1038/srep35275}.

\bibitem[\citeproctext]{ref-porter2021}
Porter, T. M., and M. Hajibabaei. 2021. {``Profile Hidden {Markov} Model Sequence Analysis Can Help Remove Putative Pseudogenes from {DNA} Barcoding and Metabarcoding Datasets.''} \emph{BMC Bioinformatics} 22 (1): 256. \url{https://doi.org/10.1186/s12859-021-04180-x}.

\bibitem[\citeproctext]{ref-rand1971}
Rand, William M. 1971. {``Objective {Criteria} for the {Evaluation} of {Clustering Methods}.''} \emph{Journal of the American Statistical Association} 66 (336): 846--50. \url{https://doi.org/10.1080/01621459.1971.10482356}.

\bibitem[\citeproctext]{ref-ratnasingham2013}
Ratnasingham, Sujeevan, and Paul D. N. Hebert. 2013. {``A {DNA-Based Registry} for {All Animal Species}: {The Barcode Index Number} ({BIN}) {System}.''} \emph{PLOS ONE} 8 (7): e66213. \url{https://doi.org/10.1371/journal.pone.0066213}.

\bibitem[\citeproctext]{ref-rognes2016}
Rognes, Torbjørn, Tomáš Flouri, Ben Nichols, Christopher Quince, and Frédéric Mahé. 2016. {``{VSEARCH}: A Versatile Open Source Tool for Metagenomics.''} \emph{PeerJ} 4 (October): e2584. \url{https://doi.org/10.7717/peerj.2584}.

\bibitem[\citeproctext]{ref-romeijn2024}
Romeijn, Luuk, Andrius Bernatavicius, and Duong Vu. 2024. {``{MycoAI}: {Fast} and Accurate Taxonomic Classification for Fungal {ITS} Sequences.''} \emph{Molecular Ecology Resources} n/a (n/a): e14006. \url{https://doi.org/10.1111/1755-0998.14006}.

\bibitem[\citeproctext]{ref-roslin2022}
Roslin, Tomas, Panu Somervuo, Mikko Pentinsaari, Paul D. N. Hebert, Jireh Agda, Petri Ahlroth, Perttu Anttonen, et al. 2022. {``A Molecular-Based Identification Resource for the Arthropods of {Finland}.''} \emph{Molecular Ecology Resources} 22 (2): 803--22. \url{https://doi.org/10.1111/1755-0998.13510}.

\bibitem[\citeproctext]{ref-schloss2009a}
Schloss, Patrick D., Sarah L. Westcott, Thomas Ryabin, Justine R. Hall, Martin Hartmann, Emily B. Hollister, Ryan A. Lesniewski, et al. 2009. {``Introducing Mothur: {Open-Source}, {Platform-Independent}, {Community-Supported Software} for {Describing} and {Comparing Microbial Communities}.''} \emph{Applied and Environmental Microbiology} 75 (23): 7537--41. \url{https://doi.org/10.1128/AEM.01541-09}.

\bibitem[\citeproctext]{ref-sibson1973}
Sibson, R. 1973. {``{SLINK}: {An} Optimally Efficient Algorithm for the Single-Link Cluster Method.''} \emph{The Computer Journal} 16 (1): 30--34. \url{https://doi.org/10.1093/comjnl/16.1.30}.

\bibitem[\citeproctext]{ref-somervuo2016}
Somervuo, Panu, Sonja Koskela, Juho Pennanen, R. Henrik Nilsson, and Otso Ovaskainen. 2016. {``Unbiased Probabilistic Taxonomic Classification for {DNA} Barcoding.''} \emph{Bioinformatics} 32 (19): 2920--27. \url{https://doi.org/10.1093/bioinformatics/btw346}.

\bibitem[\citeproctext]{ref-somervuo2017}
Somervuo, Panu, Douglas W. Yu, Charles C. Y. Xu, Yinqiu Ji, Jenni Hultman, Helena Wirta, and Otso Ovaskainen. 2017. {``Quantifying Uncertainty of Taxonomic Placement in {DNA} Barcoding and Metabarcoding.''} \emph{Methods in Ecology and Evolution} 8 (4): 398--407. \url{https://doi.org/10.1111/2041-210X.12721}.

\bibitem[\citeproctext]{ref-song2008}
Song, Hojun, Jennifer E. Buhay, Michael F. Whiting, and Keith A. Crandall. 2008. {``Many Species in One: {DNA} Barcoding Overestimates the Number of Species When Nuclear Mitochondrial Pseudogenes Are Coamplified.''} \emph{Proceedings of the National Academy of Sciences} 105 (36): 13486--91. \url{https://doi.org/10.1073/pnas.0803076105}.

\bibitem[\citeproctext]{ref-sosic2017}
Šošić, Martin, and Mile Šikić. 2017. {``Edlib: A {C}/{C}++ Library for Fast, Exact Sequence Alignment Using Edit Distance.''} \emph{Bioinformatics} 33 (9): 1394--95. \url{https://doi.org/10.1093/bioinformatics/btw753}.

\bibitem[\citeproctext]{ref-steinbach2000}
Steinbach, Michael. 2000. {``A Comparison of Document Clustering Techniques.''} Technical Report\# 00\_034/University of Minnesota.

\bibitem[\citeproctext]{ref-strehl2002}
Strehl, Alexander, and Joydeep Ghosh. 2002. {``Cluster {Ensembles} -- {A Knowledge Reuse Framework} for {Combining Multiple Partitions}.''} \emph{Journal of Machine Learning Research} 3: 583--617.

\bibitem[\citeproctext]{ref-tedersoo2024}
Tedersoo, Leho, Mahdieh S Hosseyni Moghaddam, Vladimir Mikryukov, Ali Hakimzadeh, Mohammad Bahram, R Henrik Nilsson, Iryna Yatsiuk, et al. 2024. {``{EUKARYOME}: The {rRNA} Gene Reference Database for Identification of All Eukaryotes.''} \emph{Database} 2024 (February): baae043. \url{https://doi.org/10.1093/database/baae043}.

\bibitem[\citeproctext]{ref-tedersoo2021a}
Tedersoo, Leho, Vladimir Mikryukov, Sten Anslan, Mohammad Bahram, Abdul Nasir Khalid, Adriana Corrales, Ahto Agan, et al. 2021. {``The {Global Soil Mycobiome} Consortium Dataset for Boosting Fungal Diversity Research.''} \emph{Fungal Diversity} 111 (1): 573--88. \url{https://doi.org/10.1007/s13225-021-00493-7}.

\bibitem[\citeproctext]{ref-vinh2010}
Vinh, Nguyen Xuan, Julien Epps, and James Bailey. 2010. {``Information {Theoretic Measures} for {Clusterings Comparison}: {Variants}, {Properties}, {Normalization} and {Correction} for {Chance}.''} \emph{Journal of Machine Learning Research} 11: 2837--54.

\bibitem[\citeproctext]{ref-vu2022}
Vu, Thuy, Rolf Henrik Nilsson, and Gerard Verkley. 2022. \emph{Dnabarcoder: An Open-Source Software Package for Analyzing and Predicting {DNA} Sequence Similarity Cut-Offs for Fungal Sequence Identification}. \url{https://doi.org/10.22541/au.164201896.67817672/v1}.

\bibitem[\citeproctext]{ref-wang2007}
Wang, Qiong, George M. Garrity, James M. Tiedje, and James R. Cole. 2007. {``Naïve {Bayesian Classifier} for {Rapid Assignment} of {rRNA Sequences} into the {New Bacterial Taxonomy}.''} \emph{Appl. Environ. Microbiol.} 73 (16): 5261--67. \url{https://doi.org/10.1128/AEM.00062-07}.

\bibitem[\citeproctext]{ref-watts2019}
Watts, Corinne, Andrew Dopheide, Robert Holdaway, Carina Davis, Jamie Wood, Danny Thornburrow, and Ian A Dickie. 2019. {``{DNA} Metabarcoding as a Tool for Invertebrate Community Monitoring: A Case Study Comparison with Conventional Techniques.''} \emph{Austral Entomology} 58 (3): 675--86. \url{https://doi.org/10.1111/aen.12384}.

\bibitem[\citeproctext]{ref-zito2023}
Zito, Alessandro, Tommaso Rigon, and David B. Dunson. 2023. {``Inferring Taxonomic Placement from {DNA} Barcoding Aiding in Discovery of New Taxa.''} \emph{Methods in Ecology and Evolution} 14 (2): 529--42. \url{https://doi.org/10.1111/2041-210X.14009}.

\end{CSLReferences}
\end{document}